\documentclass[twocolumn,prl,superscriptaddress,showpacs,preprintnumbers,amsmath,amssymb]{revtex4}

\usepackage{graphicx}
\usepackage{dcolumn}
\usepackage{bm}
\newcommand{\mct}{{\mbox{\scriptsize mct}}}
\newcommand{\Jacob}{J}
\newcommand{\nep}{f_{\infty}}
\newcommand{\nepself}{f_{s,\infty}}
\newcommand{\dd}{\mbox{d}}
\newcommand{\e}{\mbox{\large e}}

\newcommand{\ave}[1]{\langle {#1} \rangle}

\begin{document}
\title{Mode-Coupling Theory as a Mean-Field Description of the Glass Transition}
\author{Atsushi Ikeda}
\affiliation{Institute of Physics, University of Tsukuba, Tennodai 1-1-1, Tsukuba 305-8571, Japan}
\author{Kunimasa Miyazaki}
\affiliation{Institute of Physics, University of Tsukuba, Tennodai 1-1-1, Tsukuba 305-8571, Japan}

\date{\today}
\begin{abstract}
Mode-coupling theory (MCT) is conjectured to be a mean-field 
description of dynamics of the structural glass transition and the
 replica theory to be its thermodynamic counterpart.
However, the relationship between the two theories remains controversial and quantitative comparison is lacking. 
In this Letter, we investigate MCT for monatomic hard
sphere fluids at arbitrary dimensions above three and compare the results with replica theory.
We find grave discrepancies between the predictions of two theories. 
While MCT describes the nonergodic parameter quantitatively better than the replica theory in three dimension, 
it predicts a completely different dimension dependence of the dynamical transition point.
We find it to be due to the pathological behavior of the nonergodic parameters
derived from MCT, which exhibit negative tails in real space at high dimensions. 
\end{abstract}
 \maketitle

The nature of the glass transition remains elusive despite of decades
of discussion.      
Many theories and scenarios have been proposed to explain the drastic
slow down of dynamics of supercooled liquids but we still lack 
conclusive microscopic understanding of the phenomenon. 
Amongst various theories, 
mode-coupling theory (MCT)~\cite{Gotze2009} and replica
theory~\cite{mezard1999,Parisi2005c} are arguably the only first
principles theories.   

On the one hand, MCT describes the slow dynamics of the mildly supercooled liquids
using the static structure factor as a sole input.
It quantitatively captures the onset of the two-step relaxation of
correlation functions, the scaling properties at the intermediate time
scale (the $\beta$ relaxation), and the algebraic increase of the structural relaxation time.  
MCT, however, predicts a spurious freezing transition at  a lower
density $\varphi_{\mct}$ (or higher temperature $T_{\mct}$) than the experimentally determined
glass transition point $\varphi_g$ (or $T_g$). 
On the other hand, the replica theory is a static mean-field 
description of the glass transition~\cite{mezard1999}. 
It predicts that the fluid undergoes a thermodynamic or ``ideal''
glass transition at a higher density $\varphi_{K}$ than $\varphi_g$ 
(or lower temperature $T_K$ than $T_g$), characterized by the one-step replica symmetry breaking. 
The replica theory also predicts that the dynamical transition takes
place at $\varphi_d < \varphi_K$ (or $T_d > T_K$) where the phase space 
or energy landscape starts splitting into numerous metastable states, or
basins. 
MCT is conjectured to be the dynamical counterpart of the replica theory
and $\varphi_{\mct}$ to be identical to $\varphi_d$, 
because mathematical structure of MCT is equivalent to
the dynamical equation of the so-called $p$-spin spherical model
with $p=3$, a mean-field model for which the relation between the dynamical and ideal glass
transition is rigorously established~\cite{Kirkpatrick1987d,Castellani2005}.
According to this mean-field scenario, the absence of the dynamic transition 
at $\varphi_\mct$ in real systems is interpreted as the round-off
of the dynamic freezing by activated processes between basins in finite dimensions~\cite{Biroli2009}.  

Despite the apparent and simple parallelism with spin glasses, 
the relationship of MCT with replica theory and physical insights from the
mean-field treatments have never been fully understood.
MCT was originally derived as a generalization of kinetic theories, 
using the projection operator formalism with 
numerous uncontrolled approximations~\cite{Gotze2009}, whereas the replica theory is based on purely
thermodynamic argument developed for disordered systems. 
In this Letter, we compare the MCT and replica
theory results quantitatively, in order to clarify the relationship
between the two 
theories developed in totally different arenas of physics communities.
We especially focus on the dimension dependence of the glass transition
point and the nonergodic parameter $\nep(q)$, the plateau height of 
the density correlation function. 
To simplify the argument, we focus on the monatomic hard sphere system
in $d$-dimension, 
for which the sole system parameter is the number density $\rho=N/V$ 
or the volume fraction $\varphi=  V_d \rho $, where $V_d$ is the volume of a single hard
sphere. 
We show that MCT is more quantitative than the replica theory 
at $d=3$, which can be largely attributed to 
the lack of accurate approximation schemes in the replica  theory to
evaluate the static correlation functions of the replicated liquids.
In higher dimensions where the static correlation
functions become trivial,  discrepancies between
the two theories become catastrophic. 
MCT's dynamical transition point ($\varphi_\mct$) scales with dimension $d$ differently
from the replica counterpart ($\varphi_d$). 
This discrepancy comes from the spurious negative tails of the van
Hove correlation function, a generically positive quantity, 
that MCT predicts in high dimensions. 
This pathological negative tail is the origin of the non-Gaussian shape of $\nep(q)$ and thus
the different $d$-dependence of $\varphi_\mct$ from $\varphi_d$. 
These results shed serious doubts over the validity of MCT in higher
dimensions and call for reconsideration of MCT as a dynamic theory 
of the mean-field scenario of the glass transition.

MCT is expressed as a set of nonlinear integro-differential equations for correlation functions such
as the intermediate scattering function 
$F(q,t)$
$=$
$N^{-1}$
$\ave{\delta\rho (\vec{q},t)\delta \rho (-\vec{q},0)}$,  
where $\delta \rho (\vec{q},t)$ is the density fluctuation in reciprocal space at time $t$. 
The MCT equation for $F(q,t)$ in $d$-dimension is given by~\cite{Bayer2007b,Charbonneau2010}
\begin{eqnarray}
\Omega_q^{-2} \ddot{F}(q,t) + F(q,t) + \int^t_0 \dd s \ M(q,t-s) \dot{F}(q,s) = 0, \label{mct}
\end{eqnarray}
where $\Omega_q = \sqrt{k_BT q^2/mS(q)}$ is the phonon frequency and $S(q) =
F(q,t=0)$ is the static structure factor.
The memory function $M(q,t)$ is given by 
\begin{eqnarray}
M(q,t) = \int^{\infty}_0\!\!\!\! \dd k \int^{|q + k|}_{|q - k|}\!\!\!\! \dd p \ V(q,k,p) F(k,t) F(p,t).
\end{eqnarray}
In this expression, 
$V(q,k,p)$
$=$
$\rho$
$S(q)$
$s_{d-1}$
$\Jacob^{d-3}$
$kp$
$\{f_{+}c(k) $$+$$ f_{-} c(p) \}^2$
$/q^{d+2}$
$(4\pi)^d$
is the vertex function,
where 
$c(q) = \rho^{-1}\{ 1- 1/S(q)\}$ is the direct correlation function, $s_d$ is
the surface of $d$-dimensional unit sphere,  
$f_{\pm} = q^2 \pm(k^2 - p^2)$, 
and $\Jacob= \sqrt{4k^2p^2-(k^2-q^2+p^2)}$ is the Jacobian term.
MCT predicts a nonergodic transition at $\varphi_\mct$, beyond
which the nonergodic parameter (NEP) $\nep(q) \equiv F(q,\infty)/S(q)$
becomes non-zero.
NEP can be calculated from the long time limit of Eq.(\ref{mct}), 
\begin{eqnarray}
\frac{\nep(q)}{1-\nep(q)} = M(q, \infty).
 \label{nep} 
\end{eqnarray} 
The self-part of the intermediate scattering function 
$F_s(q, t) = \ave{ \delta \rho_s (\vec{q},t) \delta \rho_s (-\vec{q},0)}$ can be
also described by an equation similar to Eq.(\ref{mct}).
The set of MCT equations can be solved numerically using $S(q)$ as a sole
input.

\begin{figure}[htb]
\begin{center}
\includegraphics[scale=0.37]{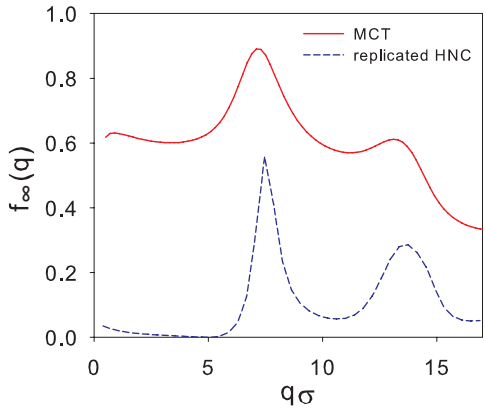}
\caption{$\nep(q)$ evaluated from MCT (solid line) and the replica theory
 (dashed line) at $\varphi_\mct$ and $\varphi_d$, respectively. 
$q$ is scaled with the diameter of a sphere $\sigma$. 
$S(q)$ from HNC closure is used for MCT calculation.
}
\label{nep_rep}
\end{center}
\vspace*{-0.4cm}
\end{figure}
On the other hand, the replica liquid theory for the structural glass transition 
has been developed by M{\'{e}}zard and Parisi~\cite{mezard1999} and recently
applied to hard-sphere fluids by Parisi and Zamponi~\cite{Parisi2005c}.
In this theory, replicated systems with a weak attractive
interaction of order $\epsilon$ between them are considered. 
The free energy of the whole system is calculated as a function of 
intra- and inter-replica correlations. 
The relevant physical observables are evaluated
by taking the limit $\epsilon \to 0$ at the end of the calculation. 
The theory predicts a thermodynamical transition characterized by the
replica symmetry breaking at which the configurational entropy vanishes.
The transition point is often identified with the Kauzmann point $\varphi_K$. 
The theory also predicts a dynamic transition 
point $\varphi_d$ $<\varphi_K$ at which the free energy splits into
numerous metastable basins. 
The system can not explore the whole phase space above $\varphi_d$ because of the 
infinite free-energy barrier which separates the basins.
In the replica interpretation, the inter-replica pair density correlation function,  
$\tilde{g}(r)$, plays the role of order parameter and is  
identified with NEP $\rho\tilde{h}(q)/S(q)=\nep(q)$, where
$\tilde{h}(q)$ is the wavevector representation of $\tilde{g}(r)-1$.
The NEP is determined from variational condition of the replicated free energy.
In analogy with the $p$-spin spherical model~\cite{Castellani2005},
it is believed that 
$\varphi_d$ should be identical to $\varphi_\mct$ and that the NEP calculated from MCT
should match that derived from the replica theory at the dynamic transition
point. 

In order to check the validity of this conjecture, we numerically compare the
results of both theories for $\varphi_{\mct}$, $\varphi_d$, and
$\nep(q)$ for $d \geq 3$.
First, we look at a $d=3$ hard sphere system. 
MCT for three-dimensional hard spheres was studied by G\"{o}tze {\it et al.}~\cite{fuchs1992,Foffi2004}.   
On the other hand, the quantitative accuracy of the replica theory sensitively depends on the
approximation scheme employed to calculate the free energy of replicated liquids.
The small cage expansion technique is known to be a good approximation
near the Kauzmann point, but it does not describe the dynamic transition
in low dimensions~\cite{mezard1999,Parisi2005c}.  
Therefore, we use another scheme, the replicated hypernetted chain
(RHNC) approximation~\cite{mezard1999}, 
the only method at present which captures the
dynamic transition in finite dimensions.
RHNC consists of a set of closure equations for both inter- and
intra-pair density correlation functions, $\tilde{g}(r)$ and ${g}(r)$,
given by 
\begin{equation}
\left\{
\begin{aligned}
&
\ln g(r) 
=  \beta v(r) 
+ \int\!\!\frac{\dd\vec{q}}{(2\pi)^d}\e^{i\vec{q}\cdot\vec{r}} 
\frac{\rho h^2(q)}{1 + \rho h(q)},
\\
&
\ln \tilde{g}(r) 
\!= \!
\int\!\! \frac{\dd\vec{q}}{(2\pi)^d}\e^{i\vec{q}\cdot\vec{r}} 
\!\!\left\{ \frac{\rho h^2(q)}{1\! +\! \rho h(q)} 
- \frac{\rho[h(q) - \tilde{h}(q)]^2}{1 + \rho [h(q)\! -\! \tilde{h}(q)]} 
\right\}.
\end{aligned} 
\right.
\label{hnc}
\end{equation}
Here, $\rho h(q) = S(q) - 1$ and $v(r)$ is the interaction potential. 
The first equation is the HNC equation of a simple liquid~\cite{hansen2006} 
and the second equation describes the inter-replica coupling. 
The dynamic transition point $\varphi_d$ is defined as the volume fraction
beyond which $\tilde{h}(q)$ becomes non-zero. 
We solve MCT equation, Eq.(\ref{nep}), and the RHNC theory,
Eq.(\ref{hnc}), for the monatomic hard sphere system.  
For MCT calculation, we employ the HNC equation to evaluate $S(q)$ in order to make the comparison consistent. 
The dynamic transition points thus obtained are $\varphi_\mct=0.523$
(which is slightly larger than 0.515 obtained from Percus-Yevick
closure~\cite{fuchs1992}) and $\varphi_d=0.612$~\cite{Velenich2006}.  
$\nep(q)$ calculated from MCT at $\varphi_\mct$ and the replica theory 
at $\varphi_d$ are shown in Figure \ref{nep_rep}.  
Quantitative difference between the
shape of the NEP from the two theories is obvious. 
Since it is well established that MCT's $\nep(q)$ agrees very well with
simulation~\cite{Foffi2004} and experimental results~\cite{vanmegen1993}, 
this discrepancy could be mainly due to poor performance of the replica theory. 
However, it is not clear whether this is attributed to 
the inherent inconsistency of MCT with replica theory or solely to  a lack
of accuracy of the RHNC approximation.

In order to give the two theories more stringent test, we 
discuss the dimension dependence of quantities near the dynamic
transition point. 
We start with $d=4-8$ and solve Eq.(\ref{nep}) to evaluate $\varphi_\mct$ and
$\nep(q)$.
An algorism by Baus and Colot~\cite{Baus1986} is used to evaluate $S(q)$. 
$\varphi_\mct$ thus obtained is listed on Table \ref{tab1} along with 
$\varphi_K$ reported in Ref.\cite{Parisi2005c}.
\begin{table}[t]
\begin{center}
\caption{Values of $\varphi_{\mct}$ and $\varphi_K$ from $d=4$ to $d=8$. 
$\varphi_K$ is from Parisi {\it et al}.~\cite{Parisi2005c}.
}
\label{table1}
\begin{tabular}{l ccccc}  \hline \hline
d &  4 &  5 & 6 & 7 & 8 \\ \hline
$\varphi_{\mct}$ ~~~~&  0.3652 ~~~& 0.2542 ~~~& 0.1736 ~~~& 0.1159 ~~~& 0.0751 \\ 
$\varphi_{K}$        &  0.4319 ~~~& 0.2894 ~~~& 0.1883 ~~~& 0.1194 ~~~& 0.0739 \\ 
\hline \hline
\end{tabular}
\label{tab1}
\end{center} 
\vspace*{-0.7cm}
\end{table}
Note that $\varphi_\mct$ is smaller than $\varphi_K$ in lower dimensions,
but the gap narrows with increasing dimension and at $d=8$ 
$\varphi_\mct$ {\it exceeds} $\varphi_K$.

Next, we study the $d$-dependence in even higher dimensions, where 
the static properties of the liquid and the replicated
liquid become insensitive to the approximation schemes.
Therefore, it is possible to check the relationship between the two theories
without obscuration from approximations for the static inputs.
In the high $d$-limit where the diagrammatic expansions of the free
energy is given by a simple function of the Mayer function 
$\e^{-\beta v(r)} - 1$, exact analytical expressions for static
correlation functions are available.   
For the hard sphere system, the direct correlation function $c(q)$ is simply given by 
$c(q)$$=$$-\left({2\pi}/{q\sigma}\right)^{d/2}$$J_{d/2}(q\sigma)$,
where $J_l(x)$ is the $l$th Bessel function of the first kind and  $\sigma$ is the diameter of a
sphere. 
Recently, accurate replica theory calculations of the free energy in high dimensions
was carried for the monatomic hard sphere system, 
using this $c(q)$ as an input and the cage expansion
method~\cite{Parisi2005c}.  
The dynamic transition point was shown to scale with $d$ as
\begin{equation}
\varphi_d = 4.8\times 2^{-d}  d
\label{eq:phid-scale}
\end{equation}
and the Kauzmann point as $\varphi_K =2^{-d}d\ln d$ 
in the high dimension limit $d\rightarrow \infty$~\cite{Parisi2005c}.
We solve the MCT equation with the same $c(q)$, 
keeping the convergence of discretization error and the numerical
accuracy of the Bessel function under control. 
\begin{figure}[htb]
\begin{center}
\includegraphics[scale=0.4]{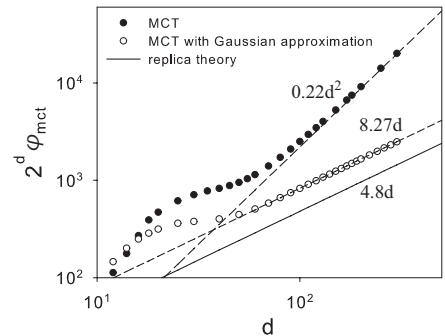}
\caption{$\varphi_\mct$ as a function of $d$.
Filled circles are the numerical solution of Eq.(\ref{nep}) 
and empty circles from MCT with Gaussian approximation, Eq.(\ref{ganep}).
Solid line is the prediction from the replica theory.}
\label{scaling}
\end{center}
\vspace*{-0.7cm}
\end{figure}
In Figure \ref{scaling}, $\varphi_\mct$ is shown as a function of $d$. 
We find that $\varphi_\mct$ scales as 
$ \varphi_\mct = 0.22 \times 2^{-d}d^2$,
in stark contrast with the replica prediction for $\varphi_d$, Eq.(\ref{eq:phid-scale}).
We also calculate the NEPs from MCT and find that their $q$-dependence
are non-Gaussian shaped in high $d$. 
In  low dimensions below $d=8$, decay of $\nep(q)$ at $q\sigma \gtrsim 10$ and $\nepself(q)$ are well
fitted by a Gaussian form, but in higher dimensions they decay faster
than Gaussian at large $q$'s (not shown).
Different $d$-dependence of $\varphi_\mct$ from $\varphi_d$ originates
from this non-Gaussianity. 
This can be shown by 
solving Eq.(\ref{nep}) assuming that 
$\nep(q)$ and $\nepself(q)$ both have a Gaussian shape, {\it i.e.}, 
$\nep(q)$, $\nepself(q)$ $\approx$ $\e^{- R q^2/{2d}}$. 
Here, we assume that $\nep(q)\approx\nepself(q)$, the so-called
Vineyard approximation~\cite{hansen2006}.
Substituting this Gaussian form in Eq.(\ref{nep}), 
we obtain a
self-consistent equation for $R$, 
\begin{eqnarray}
\frac{1}{R} = \frac{\rho s_d}{2d^2(2\pi)^d} \int_{0}^{\infty}\!\!\!\! \dd q \ q^{d+1} c^2(q) S(q) \e^{- Rq^2/d}.
\label{ganep}
\end{eqnarray}
This expression is strikingly analogous to the equation for the density field for the amorphous solid 
obtained using Gaussian approximation 
in the framework of the density functional theory~\cite{kirkpatrick1987}. 
Solving Eq.(\ref{ganep}), we find that the
equation has a finite solution for $R$ above a volume fraction
$\varphi_\mct^{(G)}$ that behaves in high $d$ 
as \footnote{The prefactor of Eq.(\ref{eq:gaphimct}) as well as (\ref{ganep}) 
differs from that in Ref.\cite{kirkpatrick1987} by a factor of two. }
\begin{equation}
 \varphi_\mct^{(G)} =  8.27 \times 2^{-d}d.
\label{eq:gaphimct}
\end{equation}
\begin{figure}[htb]
\begin{center}
\includegraphics[scale=0.35]{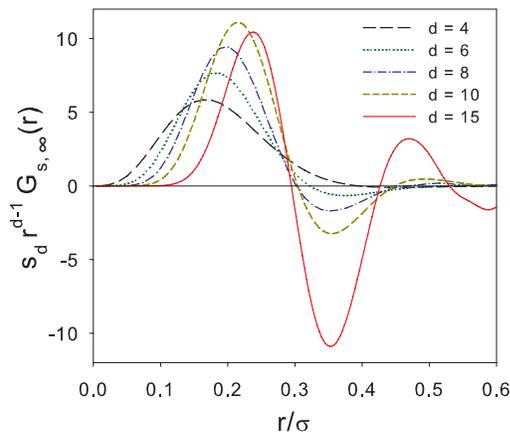}
\caption{The van Hove function $G_{s,\infty}(r)$ evaluated using MCT in several dimensions. 
From left to right, $d=4$, 6, 8, 10, 15. }
\label{bcmct_real}
\end{center}
\vspace*{-0.5cm}
\end{figure}
Though the prefactor differs, it retrieves the same $d$-dependence as 
the replica theory, which ascertains the origin of discrepancies between 
MCT and the replica theory.

This non-Gaussian shape of the NEP that MCT predicts in high dimensions
is the compelling evidence that MCT breaks down. 
Analyzing the van Hove correlation function $G_s(r, t)$, which is the real
space representation of $F_s(q,t)$, makes this breakdown clear. 
$G_s(r, t)$ is a distribution function of the distance 
for one particle to explore during the time interval $t$, 
$G_s(r, t) = \ave{\delta(r -|\vec{R}_i(t)-\vec{R}_i(0)|)}$,
where $\vec{R}_i(t)$ is the position of the $i$th particle at time $t$.
By definition, the van Hove function is a non-negative quantity. 
In Figure \ref{bcmct_real}, we plot $G_{s,\infty}(r)$
$\equiv$
$ G_s(r, t=\infty)$ derived from MCT for several dimensions. 
Following the standard convention a multiplicating factor $s_dr^{d-1}$ is used.
In high dimensions, it exhibits a negative dip whose depth becomes
larger as dimension increases.
We checked that $\nepself(q)$ retrieves the Gaussian shapes if the
negative dips are absent.

These results suggest that MCT in its present form is not consistent
with replica theory and that moreover
MCT suffers from serious deficiencies. 
The validity of the dimension dependence $\varphi_\mct \sim d^2/2^d$ 
that MCT predicts is suspicious because it originates from the
pathological behavior of the NEP.
It is noteworthy that the non-Gaussian shape of $\nep(q)$ and the
negative tails of
$G_{s,\infty}(r)$ already appear in $d=8$, below 
the upper-critical dimension $d_c=8$ of the glass transition~\cite{Biroli2007b,Biroli2006b}. 
Assuming a Gaussian shape for $\nep(q)$ in MCT recovers the 
linear dimension dependence of $2^d\varphi_\mct$ (see
Eq.(\ref{eq:gaphimct})) but the prefactor still does not match with 
the replica theory prediction. 
This fact implies that a quick remedy is unlikely to fix the problem.
In hindsight, a convincing reason to conjecture that two theories
 are related is lacking,  except for their apparent mathematical similarity 
with the $p$-spin spherical mean-field model of spin glasses. 
Deceptively similar structure between Gaussian-approximated MCT,
Eq.(\ref{ganep}), and the equation for the density profile derived in the 
mean-field analysis of the density functional theory~\cite{kirkpatrick1987} also hints that MCT is a mean-field-like theory, but 
a small yet non-negligible difference between these equations leaves us a 
nagging suspicion over MCT's validity. 
One of routes to resolve these problems is to reformulate MCT in 
a field theoretic languages in which parallels and differences 
with the  mean field theory of spin glasses 
and the dynamic liquid theory 
is highlighted. 
Efforts in this direction have suffered from series of
difficulties associated with consistencies with the
fluctuation-dissipation theorem~\cite{Miyazaki2005},
the double-counting problem of the potential interactions, and
the reconciliation between the dynamic and static liquid theories~\cite{Ikeda2010}.
Although it is not clear if the replica theory and the
density functional theory correctly describe the dynamic transition,  
at least our results clearly indicate that reconsideration and 
revision of MCT from the ground up are in order.
We conjecture, however, that prospective revisions leave
general mathematical properties of MCT equation intact. 
It is argued that MCT should be seen as a Landau theory in a sense that 
critical behavior and scaling properties 
that MCT describes near the dynamical transition point is universal~\cite{Andreanov2009b}. 
Indeed, a recent numerical study indicates that MCT works better for
critical behaviors in high dimensions~\cite{Charbonneau2010}.

\acknowledgments
 
We thank D. R. Reichman, F. Zamponi, G. Biroli, J.-P. Bouchaud, and
P. Charbonneau for stimulating discussions. This
work is supported by Grant-in-Aid for JSPS Fellows (AI),
KAKENHI; \# 21540416, (KM), and Priority Areas ``Soft Matter Physics'' 
(KM).  
\\

\noindent
{\it Note added.--}
As this manuscript was being finalized for submission, we became
aware of a paper by Schmid and Schilling~\cite{Schmid2010b}.
They have solved MCT equation for  $d\geq 10$ and
shown the same dimension dependence of $\varphi_\mct$ 
as reported in Fig.\ref{scaling} and non-Gaussian shape of $\nep(q)$.

\end{document}